\documentclass[pre,aps,floatfix,10pt,twocolumn,superscriptaddress]{revtex4-1}
\usepackage{amsmath,amssymb}
\usepackage[svgnames]{xcolor}
\usepackage{graphicx}
\usepackage{csquotes}
\usepackage{natbib}

\newcommand{\mysection}[1]{\textit{#1}.---}

\newcommand{\Ai}{\text{Ai}}
\newcommand{\Bi}{\text{Bi}}

\usepackage{acro}
\DeclareAcronym{sde}{
  short=SDE,
  long=stochastic differential equation,
}
\DeclareAcronym{ode}{
  short=ODE,
  long=ordinary differential equation,
}
\DeclareAcronym{pde}{
  short=PDE,
  long=partial differential equation,
}
\DeclareAcronym{pdf}{
  short=PDF,
  long=probability density function,
  long-format=\itshape,
}
\DeclareAcronym{cdf}{
  short=CDF,
  long=cumulative distribution function,
  long-format=\itshape,
}

\usepackage[colorlinks=true,linkcolor=DarkBlue,urlcolor=DarkBlue,citecolor=FireBrick]{hyperref}

\begin{document}

\title{Predictability of escape for a stochastic saddle-node bifurcation: \\ when rare events are typical}
\author{Corentin Herbert}
\email{corentin.herbert@ens-lyon.fr}
\affiliation{Univ Lyon, ENS de Lyon, Univ Claude Bernard, CNRS, Laboratoire de Physique, F-69342 Lyon, France}
\author{Freddy Bouchet}
\affiliation{Univ Lyon, ENS de Lyon, Univ Claude Bernard, CNRS, Laboratoire de Physique, F-69342 Lyon, France}

\begin{abstract}
  Transitions between multiple stable states of nonlinear systems are ubiquitous in physics, chemistry, and beyond.
  Two types of behaviors are usually seen as mutually exclusive: unpredictable noise-induced transitions and predictable bifurcations of the underlying vector field.
  Here, we report a new situation, corresponding to a fluctuating system approaching a bifurcation, where both effects collaborate.
  We show that the problem can be reduced to a single control parameter governing the competition between deterministic and stochastic effects.
  Two asymptotic regimes are identified: when the control parameter is small (e.g. small noise), deviations from the deterministic case are well described by the Freidlin-Wentzell theory.
  In particular, escapes over the potential barrier are very rare events.
  When the parameter is large (e.g. large noise), such events become typical.
  Unlike pure noise-induced transitions, the distribution of the escape time is peaked around a value which is asymptotically predicted by an adiabatic approximation.
  We show that the two regimes are characterized by qualitatively different reacting trajectories, with algebraic and exponential divergence, respectively.
\end{abstract}


\maketitle

\begin{figure}
  \includegraphics[width=\linewidth]{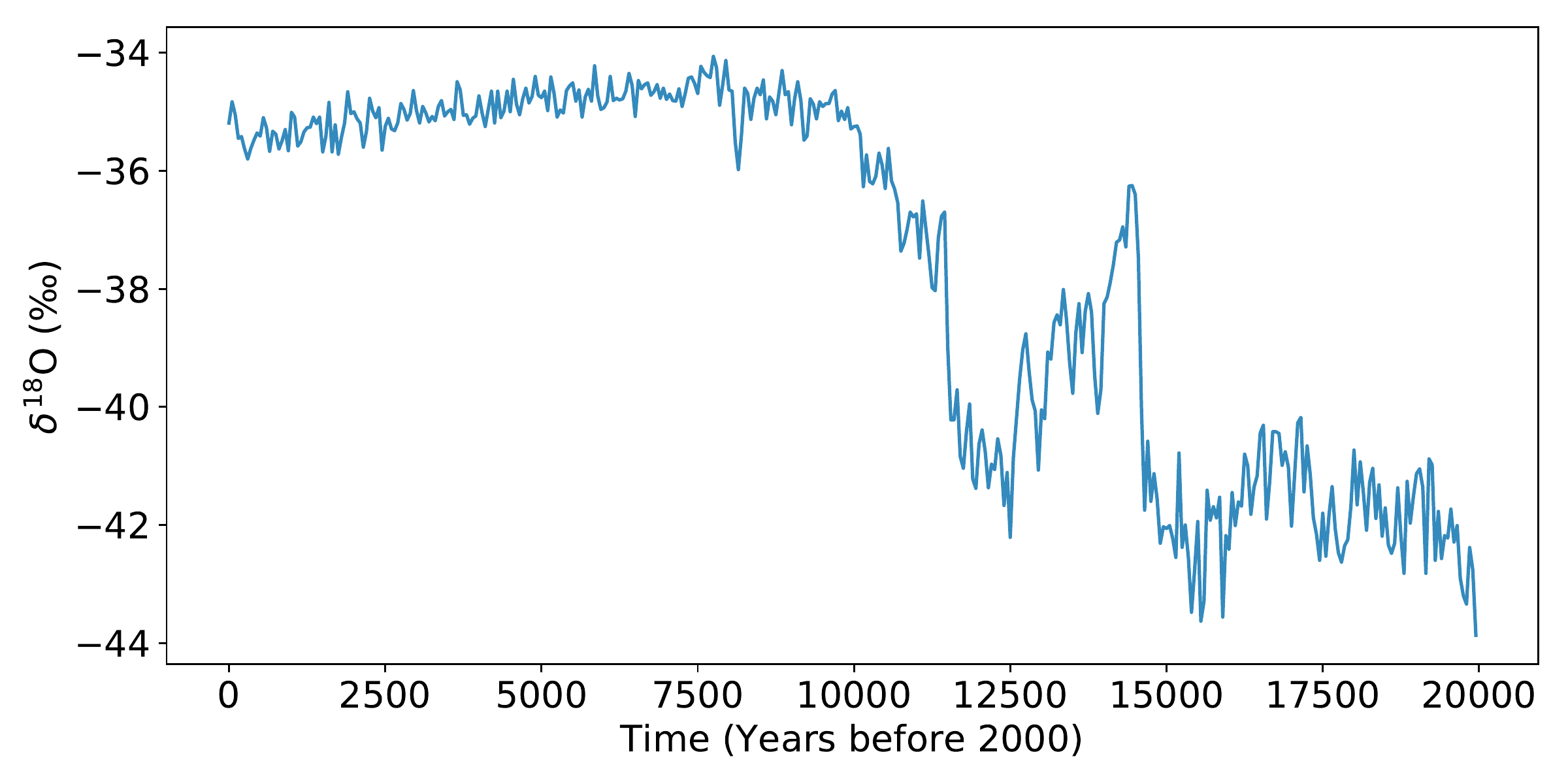}
  \caption{\label{fig:cycles} Paleoclimatic oxygen isotopic record (a proxy for temperature) from NGRIP ice core~\cite{NGRIP2004}, Greenland, zoomed on the last termination. Do such trajectories have universal properties?}
\end{figure}

% Abrupt transitions
Abrupt transitions between distinct statistically steady states are generic features of complex dynamical systems.
Although usually very rare, such events are extremely important because the qualitative behavior of the system may change radically.
For instance, abrupt and dramatic transitions are frequently encountered in climate dynamics, from the global Neoproterozoic glaciations (\emph{snowball Earth} events)~\cite{Pierrehumbert2011}, to glacial-interglacial cycles (see Fig.~\ref{fig:cycles}) of the Pleistocene~\cite{Paillard1998,Huybers2005,Crucifix2013}, to the rapid Dansgaard-Oeschger events~\cite{Dansgaard1993,Ganopolski2002,Ditlevsen2007}.
The timing and amplitude of the transitions rule out the possibility of a linear response to an external forcing.
% Bifurcations
Like in many physical systems, such as bistable lasers~\cite{Jung1990} or ferromagnets~\cite{Rao1990,Lo1990}, these transitions may instead be due to a parameter crossing a critical threshold, resulting in structural modifications in the internal dynamics, i.e. a bifurcation.
Indeed, mechanisms accounting for multistability and hysteresis in the climate system have been evidenced in a wide variety of contexts~\cite{Rahmstorf2002,Dijkstra2005,Eisenman2009,Rose2013}.
% Noise-induced scenario
On the other hand, intrinsic variability, represented as noise acting on the variable of interest, may be responsible for spontaneous transitions on very long timescales, in much the same way as diffusion-controlled chemical reactions~\cite{Arrhenius1889,Eyring1935,Kramers1940,Calef1983}, tunneling in quantum mechanical systems~\cite{Bourgin1929,Wigner1932} or transitions in hydrodynamic~\cite{Ravelet2004,Bouchet2009} or magnetohydrodynamic~\cite{Berhanu2007} turbulence.
% more examples: Josephson junctions, etc
The problem of noise-activated transitions in a time-varying potential is therefore of broad interest, with many practical applications across various fields of physics.
For instance, ramping up or modulating periodically a bifurcation parameter is a widely used technique to probe small systems subjected to noise --- e.g. thermal noise in Josephson junctions~\cite{Kurkijarvi1972} or, more generally, out-of-equilibrium nonlinear oscillators~\cite{Dykman1980,Dykman2012}.

Motivated by the possibility to predict the approach of a tipping point, many earlier studies have focused on \emph{early-warnings}, i.e. features of a time-series which change before the transition occurs.
In that framework, deterministic bifurcations are announced by phenomena such as increasing autocorrelation or variance, which are absent in noise-induced transitions~\cite{Scheffer2009,Ditlevsen2010,Thompson2011}.
Here, we adopt a different point of view, and study the universal statistical and dynamical features of transitions occurring under the joint effect of loss of stability and stochastic forcing.
In the classical noise-induced case, transitions are completely random (they follow a Poisson distribution), the reaction rate satisfies the Arrhenius law~\cite{Hanggi1990}, and typical reacting trajectories follow the optimal path minimizing the action, as predicted by Freidlin-Wentzell theory~\cite{FreidlinWentzellBook}.
On the other hand, the transition is completely governed by the deterministic behavior in the bifurcation scenario.
In this paper, we study when and how the transition occurs under a sweeping of the bifurcation parameter in time, in the presence of noise.
An important motivation is to understand how much can be learned about the transition from observations of trajectories such as those represented in Fig.~\ref{fig:cycles}.
What is the parameter governing the competition between deterministic and stochastic behavior?
Can we distinguish trajectories corresponding to these two regimes?
We show that the transition time in a system approaching loss of stability is controlled by a single parameter and is always more predictable than the static noise-induced case.
When the noise is small, the behavior is close to deterministic, the particle escapes slightly after the bifurcation and follows a universal trajectory with algebraic divergence.
In the large noise regime, the escape time is determined by a balance between deterministic (lowering of the potential barrier) and stochastic effects, similarly to \emph{stochastic resonance}~\cite{Benzi1981,Gammaitoni1998}.
We show that the probability distribution of the escape time reaches a peak well before the bifurcation time, and can be predicted by an adiabatic approximation, corresponding to an Eyring-Kramers regime.
Typical reacting trajectories leave the attractor in an exponential manner, and they show no imprint of the saddle-node, unlike the standard time-independent case.

\begin{figure}
  \includegraphics[width=\linewidth]{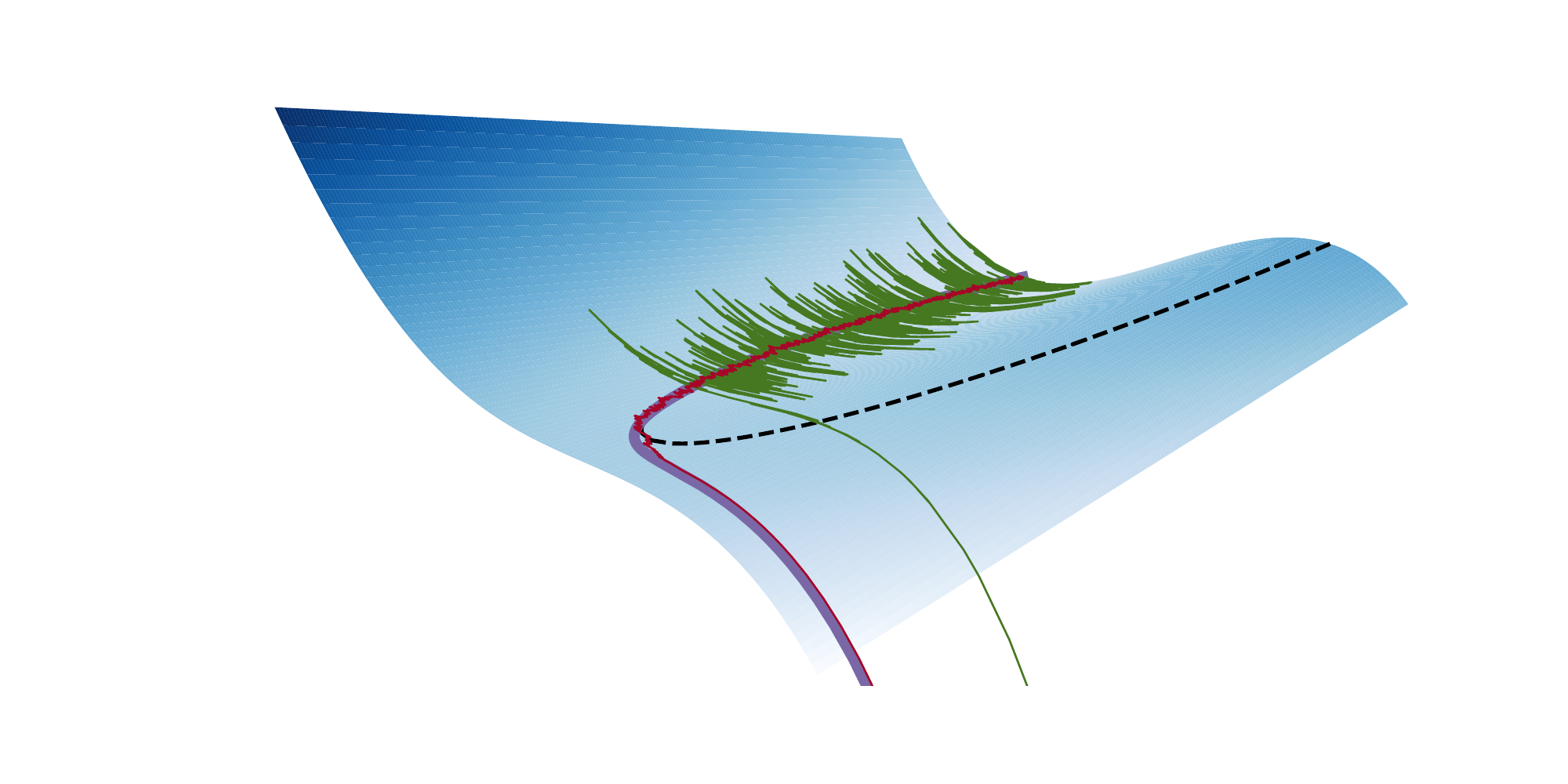}
  \caption{\label{fig:potential} (Color online) Potential $V(x,t)$ undergoing a saddle-node bifurcation, and sample trajectories for the stochastic process described by Eq.~\ref{eq:sde}: deterministic attractor (thick purple curve), small-noise trajectory ($\epsilon=0.1$, red) exiting near the bifurcation, and escape over the potential barrier ($\epsilon=10$, green). The dashed curve indicates the position of the saddle-point $x_+(t)$.}
\end{figure}

\mysection{The model}
% Non-dimensionalization and reduction to a single control parameter
Let us consider an overdamped Langevin particle in a time-dependent potential $V(x,t)$, undergoing a saddle-node bifurcation at $t=0$.
The system is described by the stochastic differential equation:
\begin{equation}
  dx_t = -\partial_x V(x_t,t)dt+\sqrt{2\sigma}dW_t, \label{eq:sde}
\end{equation}
where $W_t$ is the standard Brownian motion.
The most simple such potential has the form $V(x,t)=-a^3x^3/3-a\omega t x$, where the spatial scale $a$ and the time-dependent bifurcation parameter $\omega t$ determine the height of the potential barrier $\Delta V = 4 (-\omega t)^{3/2}/3$ and the width of the potential well $\sqrt{-3\omega t/a^{1/3}}$.
By a proper choice of units, all the relevant parameters (geometry of the potential, speed of approach of the bifurcation, and noise amplitude) can be absorbed into a single, non-dimensional parameter $\epsilon=\sigma/(a^{2/3}\sqrt{\omega})$.
With the rescaled variables, the potential now reads $V(x,t)=-x^3/3-tx$, and the stochastic differential equation is the normal form for the saddle-node bifurcation (with time-dependent bifurcation parameter) perturbed by noise: $dx_t = (x_t^2+t)dt+\sqrt{2\epsilon}dW_t$.
We should keep in mind that the universality of the saddle-node normal form is only valid close to the bifurcation.
Therefore, we expect our results to be universal for slow enough bifurcation parameter drift for arbitrary potentials.
We shall denote by $x_\pm(t)=\pm\sqrt{-t}$ the fixed points for the stationary problem, which exist for $t<0$ only.
The particle, initially lying in the stable state ($x_0=x_-(t_0)$),  may escape over the potential barrier under the influence of noise, or simply follow the deterministic dynamics and escape after the potential barrier has been removed by the bifurcation (see Fig.~\ref{fig:potential})~\cite{Berglund2002,BerglundGenzBook,Kuehn2011,Miller2012}.
The single control parameter $\epsilon$ governs the competition between stochastic and deterministic effects.
It decreases with the speed of the bifurcation and the potential stiffness, and increases with noise.

% Defining the first-passage time
To give a precise meaning to the notion of escape, we shall compute the probability distribution of the \emph{first passage time}, defined by
\begin{equation}
  \tau_M(x_0,t_0) = \inf \{ t\geq t_0 | x_t \geq M \}. \label{eq:fpt}
\end{equation}
Given the shape of the potential, the results do not depend on $M$ for $M$ large enough.
For homogeneous Markov processes, a closed set of equations for the moments $\mathbb{E}[\tau_M^n]$ may be obtained, which leads to an explicit quadrature formula for the mean first-passage time for a 1D system~\cite{GardinerBook,RiskenBook}.
Since the stochastic process defined by Eq.~\ref{eq:sde} is not time-homogeneous, these theoretical results do not apply here.
We will discuss the behavior of the random variable $\tau_M$ using numerical results obtained with standard Monte-Carlo simulations and numerical solutions of the Fokker-Planck equation associated to Eq.~\ref{eq:sde}, as well as theoretical arguments in the two limiting regimes $\epsilon \ll 1$ and $\epsilon \gg 1$.

\mysection{Deterministic and small-noise behavior}
In the deterministic case ($\epsilon=0$), we have the \emph{dynamical saddle-node} bifurcation, for which an analytical solution for the trajectory $x(t;x_0,t_0)$ with initial conditions $x_0,t_0$ can be found in terms of Airy functions.
In particular, the attractor $\bar{x}(t)=\lim_{t_0 \to -\infty} x(t;x_-(t_0),t_0)$ simply reads $\bar{x}(t)=\Ai'(-t)/\Ai(-t)$.
When $-t$ is large, it follows the stationary solution $x_-(t)=-\sqrt{-t}$.
At a time of order one before the bifurcation ($t=0$), the trajectory detaches, and diverges to infinity after the bifurcation (see Fig.~\ref{fig:potential}).
The singularity occurs at a time $t_\star \approx 2.33811$, which is the opposite of the largest root of the Airy function $\Ai$.
The divergence is algebraic: $\bar{x}(t) \sim (t_\star-t)^{-1}$, and the deterministic first-passage time is easily related to the singularity: $\bar{\tau}_M=t_\star-1/M+o(1/M)$.

When the noise amplitude $\epsilon$ is small, escapes over the potential barrier have so low (albeit non-vanishing) probability that the behavior of the system is dominated by escapes after the bifurcation occurs~\cite{Berglund2002,BerglundGenzBook,Kuehn2011,Miller2012}.
This regime is close to the deterministic behavior.

\begin{figure}
  \includegraphics[width=\linewidth]{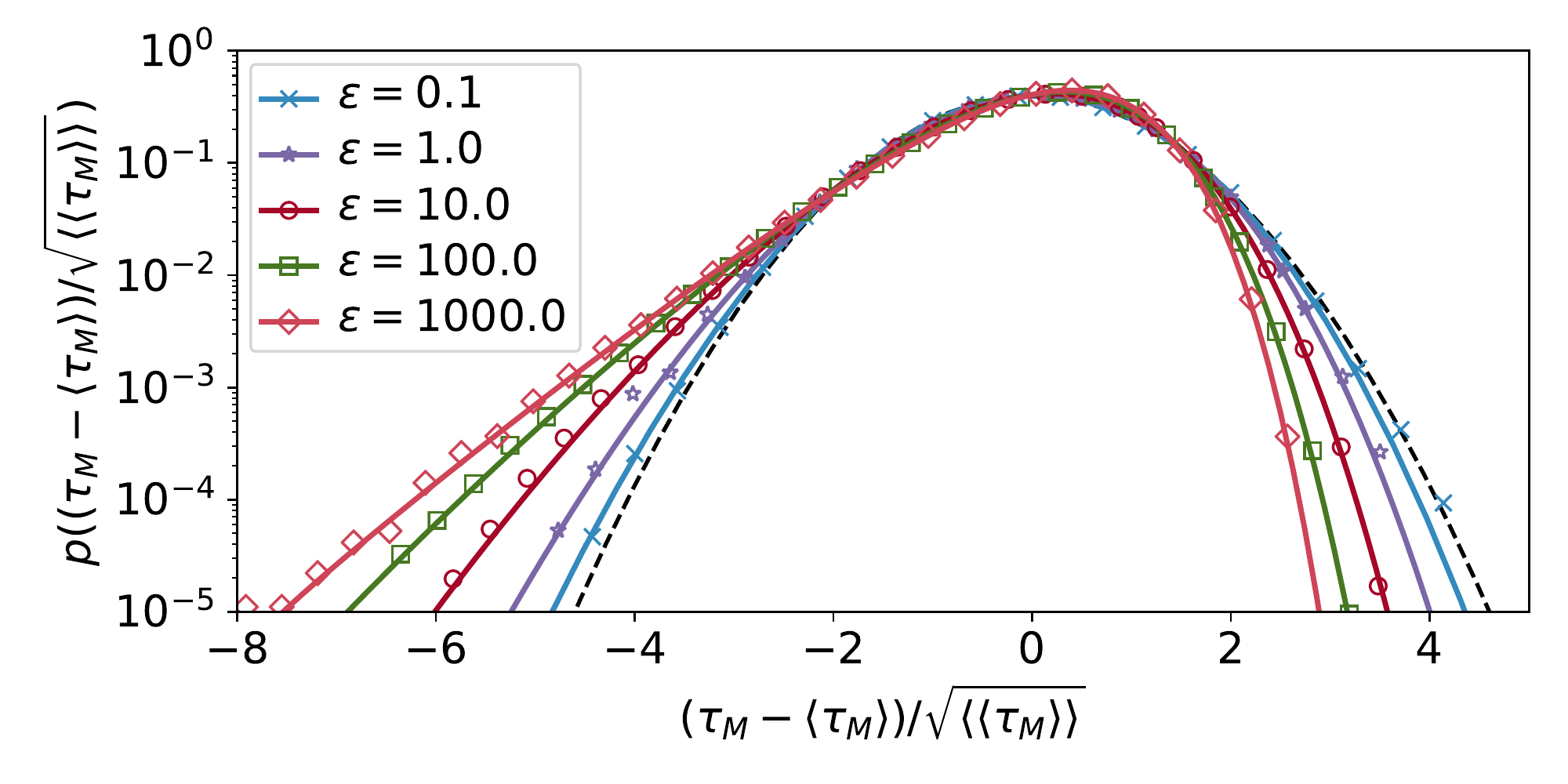}
  \caption{\label{fig:fptpdfnormalized} (Color online) Standardized \ac{pdf} of the first-passage time $\tau_M$ ($M=20$), for different values of $\epsilon$, obtained by Monte-Carlo simulations (symbols) and numerical solution of the Fokker-Planck equation (lines). The dashed black line is the standard normal distribution.}
\end{figure}

\begin{figure}
  \includegraphics[width=\linewidth]{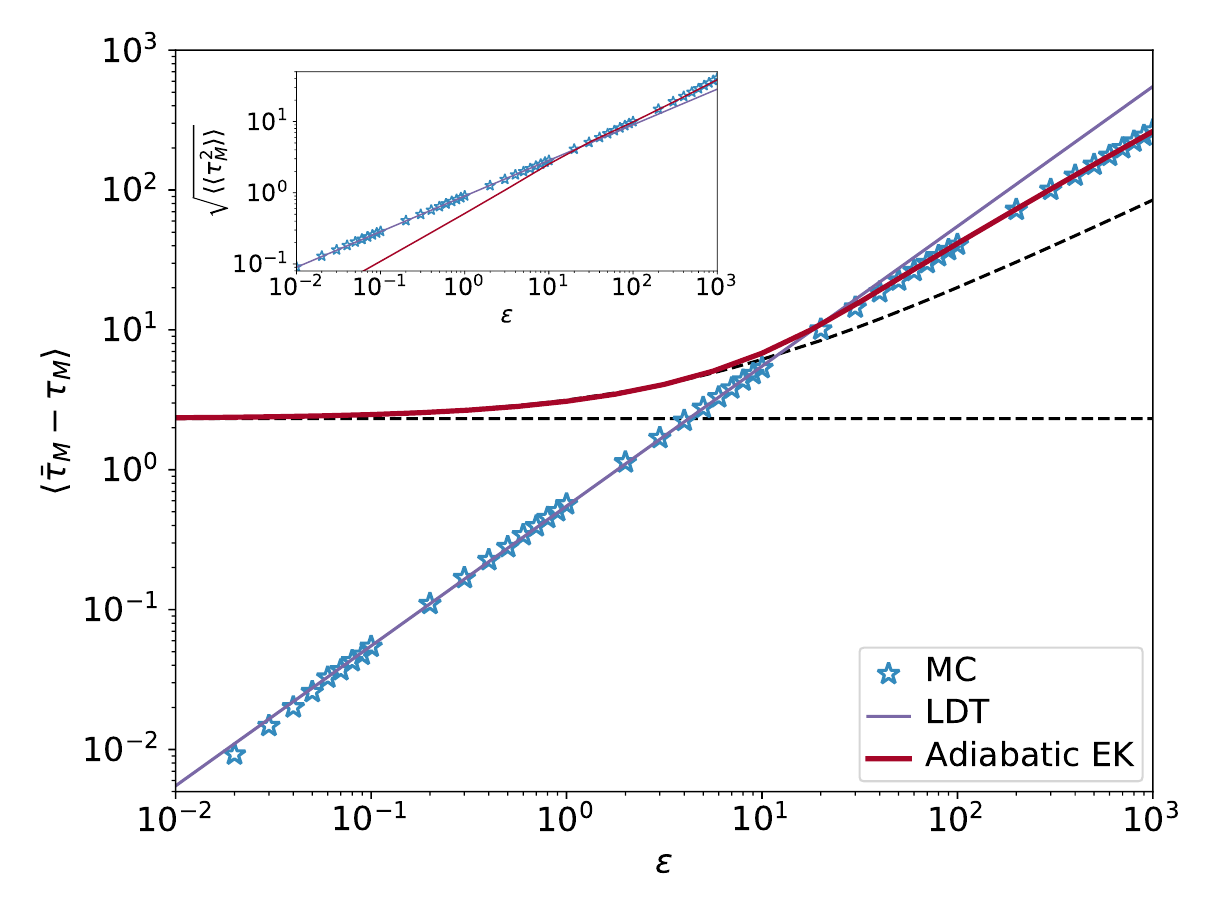}
  \caption{\label{fig:fptavg} (Color online) Mean first-passage time $\langle \tau_M \rangle$ ($M=20$), subtracted from the deterministic value $\bar{\tau}_M$, obtained by Monte-Carlo simulations (stars), by large deviation theory (thin purple curve), and by the adiabatic Eyring-Kramers ansatz (thick red curve). The horizontal dashed line corresponds to the bifurcation time $t=0$. The other dashed curve corresponds to the times such that $\Delta V(t)=\epsilon$. Inset: standard deviation.}
\end{figure}

The \ac{pdf} of the first-passage time $\tau_M$, computed numerically, is shown in Fig.~\ref{fig:fptpdfnormalized}.
When $\epsilon$ is small, the \ac{pdf} is close to Gaussian.
As $\epsilon$ increases, the \ac{pdf} becomes more and more skewed, and heavy tails develop on the left, due to the presence of early exits.
This can be interpreted in the framework of large deviation theory~\cite{FreidlinWentzellBook}.
Let us assume that the \ac{pdf} of $\tau_M$ satisfies a large deviation principle: $p_M(\tau) \asymp e^{-I_M(\tau)/\epsilon}$ for $\epsilon \to 0$.
If the rate function $I_M$ possesses a single zero $\tau_\star$ (also a global minimum), then the Gaussian behavior of $\tau_M$ corresponds to a quadratic approximation around $\tau_\star$.
Besides, the integral defining the mean first-passage time $\mathbb{E}[\tau_M]$ can be evaluated with a saddle-point approximation.
At first order, we expect the deterministic first-passage time $\bar{\tau}_M$, and the error associated with the approximation is linear in $\epsilon$~\cite{ErdelyiBook}: $\mathbb{E}[\tau_M]=\bar{\tau}_M (1+O(\epsilon))$.
Based on similar considerations, the asymptotic behavior of the standard deviation is expected to be $\sqrt{\mathbb{E}[[\tau_M^2]]}\sim \sqrt{\epsilon}$.
These provide a good fit of numerical results, as shown in Fig.~\ref{fig:fptavg}.

In the limit $M\to \infty$, Ref.~\cite{Miller2012} provides an exact result which shows that the random variable $\tau_{\infty}$ (the time at which the trajectory becomes unbounded) satisfies a large deviation principle, with rate function $I_\infty(\tau)=\Ai^2(-\tau)/\lbrack4\pi^2\Bi^2(-\tau)\int_{-\infty}^T \Ai^4(-t)dt\rbrack$, when $\tau$ lies in the range between the first two zeros of $\Bi$.
In this interval, the only root of $I_\infty$ is $t_\star$, and the above reasoning applies, with $\mathbb{E}[\tau_\infty] \to t_\star$.

More generally, the large deviation property for $p_M$ can be understood at a formal level by writing the probability of fluctuations within the path integral formalism, following the pioneering work of Onsager and Machlup~\cite{Onsager1953,*Machlup1953}.
In the $\epsilon \ll 1$ regime, the probability of a path $x(t)$ satisfies $P[x] \asymp e^{-\mathcal{A}[x]/\epsilon}$, introducing the Freidlin-Wentzell action functional $\mathcal{A}[x]=\frac{1}{4}\int dt (\dot{x}+V'(x,t))^2$~\cite{FreidlinWentzellBook}.
This probability distribution is dominated by the deterministic attractor $\bar{x}(t)$, for which $\mathcal{A}[\bar{x}]=0$.
By contraction, the probability to reach $M$ at time $\tau$ is dominated by another action minimizer $x^\star_{M,\tau}(t)$ such that $\mathcal{A}[x^\star_{M,\tau}]= A_M(\tau) \equiv\inf_x\{\mathcal{A}[x] | x(t_0)=x_0, x(\tau)=M\}$.
As we shall see below, such optimal paths follow the deterministic attractor as long as possible, at no cost in the action functional, then detach from the deterministic trajectory in a monotonously increasing manner.
The monotonicity of the action minimizers can be proved directly by writing the Hamilton equations associated to the variational problem:
\begin{equation}
  \dot{x} = x^2+t+2p, \quad \dot{p} = -2xp.
\end{equation}
As a consequence, there is no optimal path which reaches $M$ more than once, because of the cost in the action functional associated with climbing the gradient of the potential.
Hence, the trajectories $x^\star_{M,\tau}(t)$ dominate the \ac{pdf} of the first-passage time $\tau_M$, which satisfies a large deviation principle, with rate function $I_M(\tau)=A_M(\tau)$.
Then, the Gaussian behavior of $\tau_M$ corresponds to a quadratic approximation of $A_M(\tau)$ around its minimum $\bar{\tau}_M$.

In fact, although the \ac{pdf} of $\tau_M$ already exhibits substantial deviation from Gaussianity for $\epsilon>0.1$ (see Fig.~\ref{fig:fptpdfnormalized}), the above approach describes accurately the first two moments up to order one values of $\epsilon$.
For these low-order statistics, a sharp transition between the asymptotic regimes $\epsilon \ll 1$ and $\epsilon \gg 1$ occurs near $\epsilon=20$ (see Fig.~\ref{fig:fptavg}).
On average, the transition always happens before the deterministic case.
It happens before the bifurcation for $\epsilon>\epsilon_c \approx 4$, and after for $\epsilon<\epsilon_c$.
The discrepancies seen in Fig.~\ref{fig:fptavg} for the mean first-passage time at very small $\epsilon$ are due to numerical errors.

\begin{figure}
  \includegraphics[width=\linewidth]{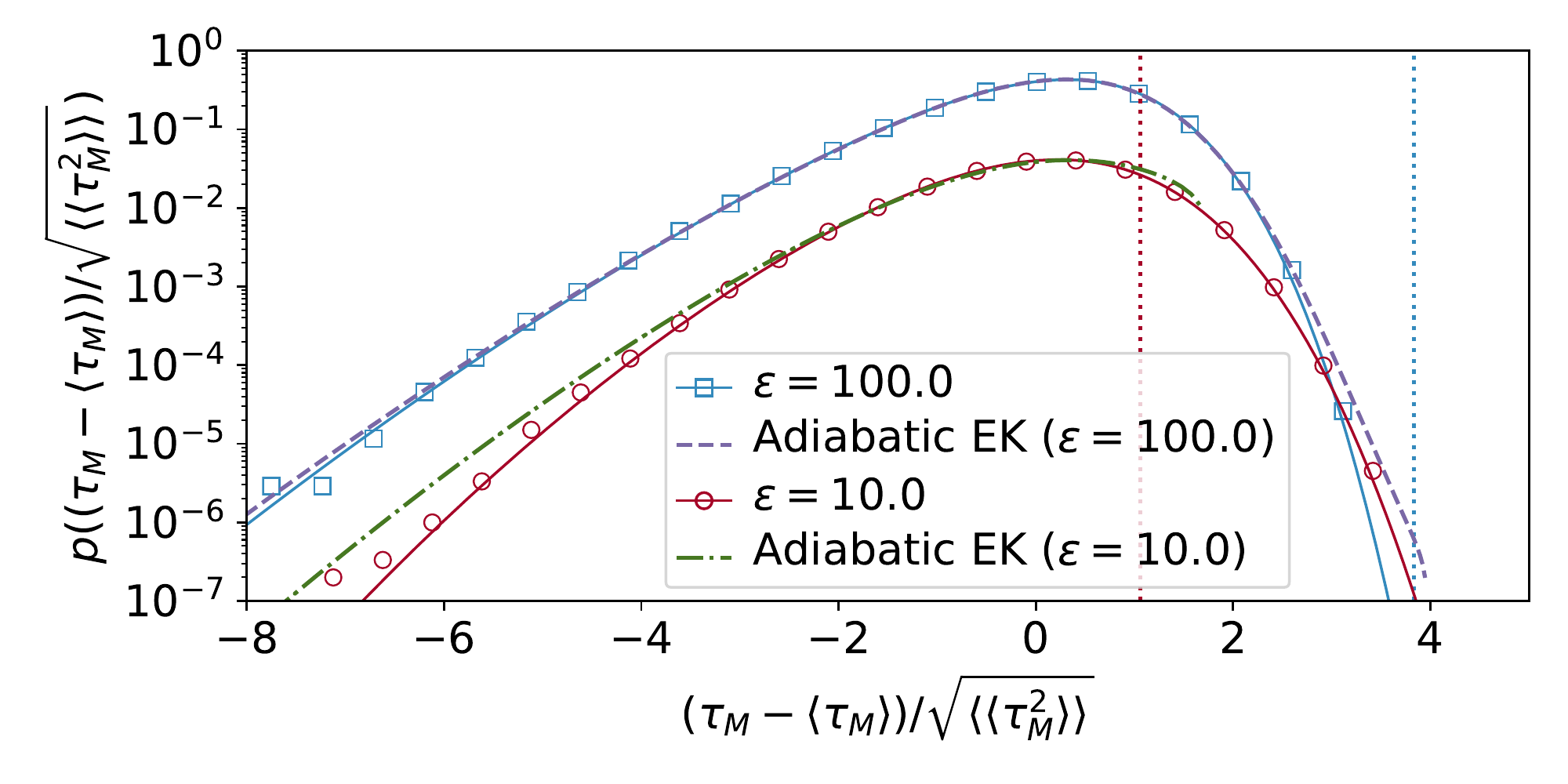}
  \caption{\label{fig:fptpdf} (Color online) \ac{pdf} of the first passage time in the large noise regime, computed with Monte-Carlo simulations (symbols), numerical solution of the Fokker-Planck equation (solid curves), compared with the theoretical curve obtained with the adiabatic approximation and the Eyring-Kramers ansatz (dashed curves), for $\epsilon=100$ and $\epsilon=10$. The vertical dotted lines indicate escapes occurring at the bifurcation time ($\tau_M=0$). The $\epsilon=10$ curves are shifted downwards by a factor 10 for clarity.}
\end{figure}

\mysection{Adiabatic approximation in the large-noise regime}
When the noise amplitude is large, the range of times for which the escape rate is not too small is long enough for those events to dominate the distribution of the first-passage time.
In this regime, escapes over the potential barrier, which are usually rare events, become the typical events.
However, that range is also short enough for the distribution of the first-passage time to be peaked around a given value (Fig.~\ref{fig:fptpdf}), determined by the competition between stochastic and deterministic effects.
This is very different from the classical Kramers problem, for which the first-passage time is distributed according to an exponential law~\cite{Hanggi1990}.

Because the relaxation time scale is much smaller than the scale at which the potential evolves, this case can be treated with an adiabatic approximation.
We introduce the transition probability $P(x,t|x_0,t_0)$, which satisfies the (forward) Fokker-Planck equation with initial condition $P(x,t_0|x_0,t_0)=\delta(x-x_0)$.
With reflecting boundary condition on the left and absorbing boundary condition at a fixed value $M>x_+(t_0)$, $G(x_0,t_0;M,t)=\int_{-\infty}^M dx P(x,t|x_0,t_0)$ is the probability that a particle initially at $x_0$ has not reached $M$ at time $t$.
In other words, $\text{Prob}(\tau_M>t)=G(x_0,t_0;M,t)$.
$G$ always satisfies a backward Fokker-Planck equation.
For homogeneous Markov processes, because $\partial_t G=-\partial_{t_0} G$, this partial differential equation allows to compute explicitly the moments of the first-passage time~\cite{GardinerBook}.
Besides, when the potential barrier height $\Delta V$ is large ($\Delta V \gg \epsilon$), transition times form a Poisson process with transition rate given by the Eyring-Kramers formula: $\lambda=\sqrt{V''(x_a)V''(x_s)}/(2\pi)e^{-\Delta V/\epsilon}$, where $x_a$ is the position of the attractor and $x_s$ that of the saddle point~\cite{Berglund2013}.
Here, since the potential variations are adiabatic, the transition rate at each time $t$ is well approximated by the Eyring-Kramers formula for the \enquote{frozen} potential at fixed $t$: $\partial_t G(x_0,t_0;M,t)=-\lambda_{EK}(t) G(x_0,t_0;M,t)$, with $\lambda_{EK}(t)=\frac{\sqrt{-t}}{\pi}\exp(-\Delta V(t)/\epsilon)$, where $\Delta V(t) = 4(-t)^{3/2}/3$.
This formula is expected to be valid for times $t \ll -\epsilon^{2/3}$, and initial conditions close to the attractor.
An explicit formula is obtained for the \ac{pdf} of the first-passage time in this regime:
\begin{equation}
  \mathbb{E}_{EK}[\delta(\tau_M-t)] = \frac{\sqrt{-t}}{\pi}e^{-\frac{4(-t)^{3/2}}{(3\epsilon)}}\exp\left\lbrack -\frac{\epsilon}{2\pi}e^{-\frac{4(-t)^{3/2}}{(3\epsilon)}}\right\rbrack. \label{ekpdfeq}
\end{equation}
We show in Fig.~\ref{fig:fptpdf} that this approximation indeed provides a very good fit of the numerically computed \ac{pdf} of the first-passage time $\tau_M$ when $\epsilon$ is large enough (here $\epsilon=100$).
From Eq.~\ref{ekpdfeq}, we deduce the asymptotic behavior of the moments of the first-passage time: when $\epsilon \to +\infty$,
\begin{equation}
  \mathbb{E}_{EK}[\tau_M^n] \sim \frac{(-1)^n}{2\pi} \left(\frac{3\epsilon \ln \epsilon}{4}\right)^{2n/3}. \label{ekmomentseq}
\end{equation}
The mean first-passage time and its standard deviation are shown in Fig.~\ref{fig:fptavg}; again, the theoretical result fits very well the numerical simulations above a critical $\epsilon$ approximately equal to 20.
Besides, for the adiabatic approximation to be self-consistent, we need $\mathbb{E}_{EK}[\tau_M] \ll -\epsilon^{2/3}$.
This condition is asymptotically verified, but this is only due to the logarithmic corrections in Eq.~\ref{ekmomentseq}.
Hence, the adiabatic approximation converges relatively slowly.
This explains why the theoretical result is not very accurate for $\epsilon=10$ for instance (see Fig.~\ref{fig:fptpdf}).
For such moderate values of the control parameter, the approximation slightly overestimates early escapes, and makes a dramatic error on escapes occurring later than the average time.
Indeed, for such regimes, escapes after the bifurcation occurs, which make no sense in the Eyring-Kramers approximation, are already relatively probable events (about one or two standard deviations away from the mean).
Although such events are still unaccounted for at larger $\epsilon$, they are then so improbable that it does not hamper the accuracy of the approximation for low-order moments.

\mysection{Predictability of the reacting trajectory}
Now, we consider the statistics of the escape dynamics.
\begin{figure}
  \includegraphics[width=\linewidth]{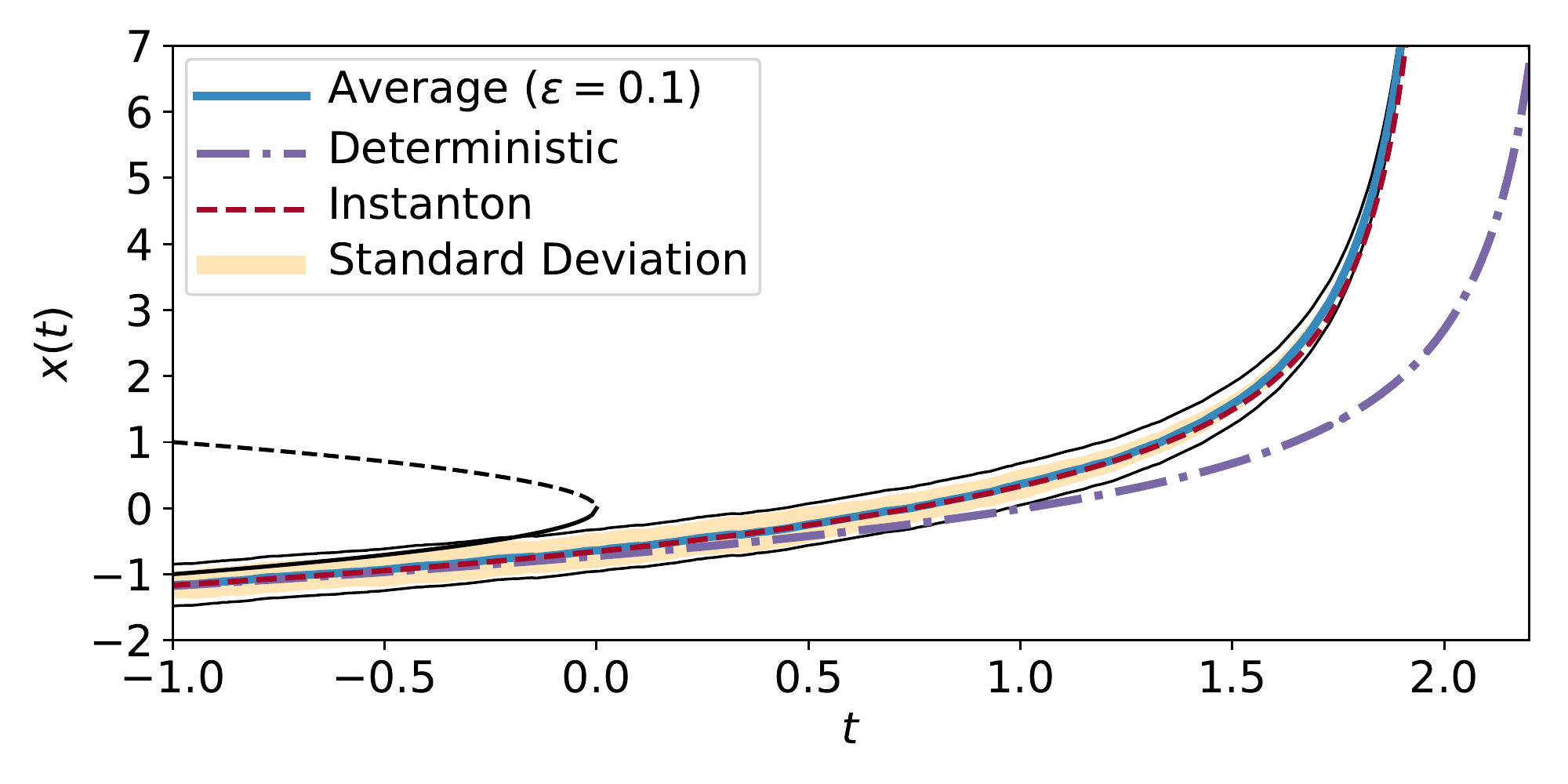}
  \caption{\label{fig:condtrajsmall} (Color online) Average trajectory $\mathbb{E}[x_t|\tau_M]$ (solid blue curve) and standard deviation (shading) for the trajectories conditioned on the first-passage time ($\tau_M=2$), for $M=20$ and $\epsilon=0.1$, compared with the instanton (dashed red). The deterministic attractor $\bar{x}(t)$ (dash-dotted purple curve) has a different $\tau_M$ but the same shape as reacting trajectories. Thin black lines indicate an error of order $\sqrt{\epsilon}$.}
\end{figure}
For small $\epsilon$, the particle typically escapes after the bifurcation, and the dynamics is then essentially deterministic.
Hence, all the reacting trajectories have the same shape as the deterministic attractor, even though escape typically occurs slightly before.
This can be illustrated by conditioning trajectories on the first-passage time.
Fig.~\ref{fig:condtrajsmall} shows that, when conditioning on a typical value for the first-passage time (less than one standard deviation away from the mean), apart from fluctuations of order $\sqrt{\epsilon}$, the reacting trajectories remain close to a trajectory with the same shape as the deterministic attractor $\bar{x}(t)$.
That trajectory can be predicted as an \emph{instanton}: it is a minimizer of the action $\mathcal{A}[x]$ with fixed initial and final points.
In particular, it has an algebraic divergence of the form $x(t)\sim (t_\star'-t)^{-1}$, where $t_\star'=\tau_M+1/M$ for trajectories conditioned on $\tau_M$.
Even rare transitions look similar to the deterministic attractor and are hardly distinguishable.

\begin{figure}
  \includegraphics[width=\linewidth]{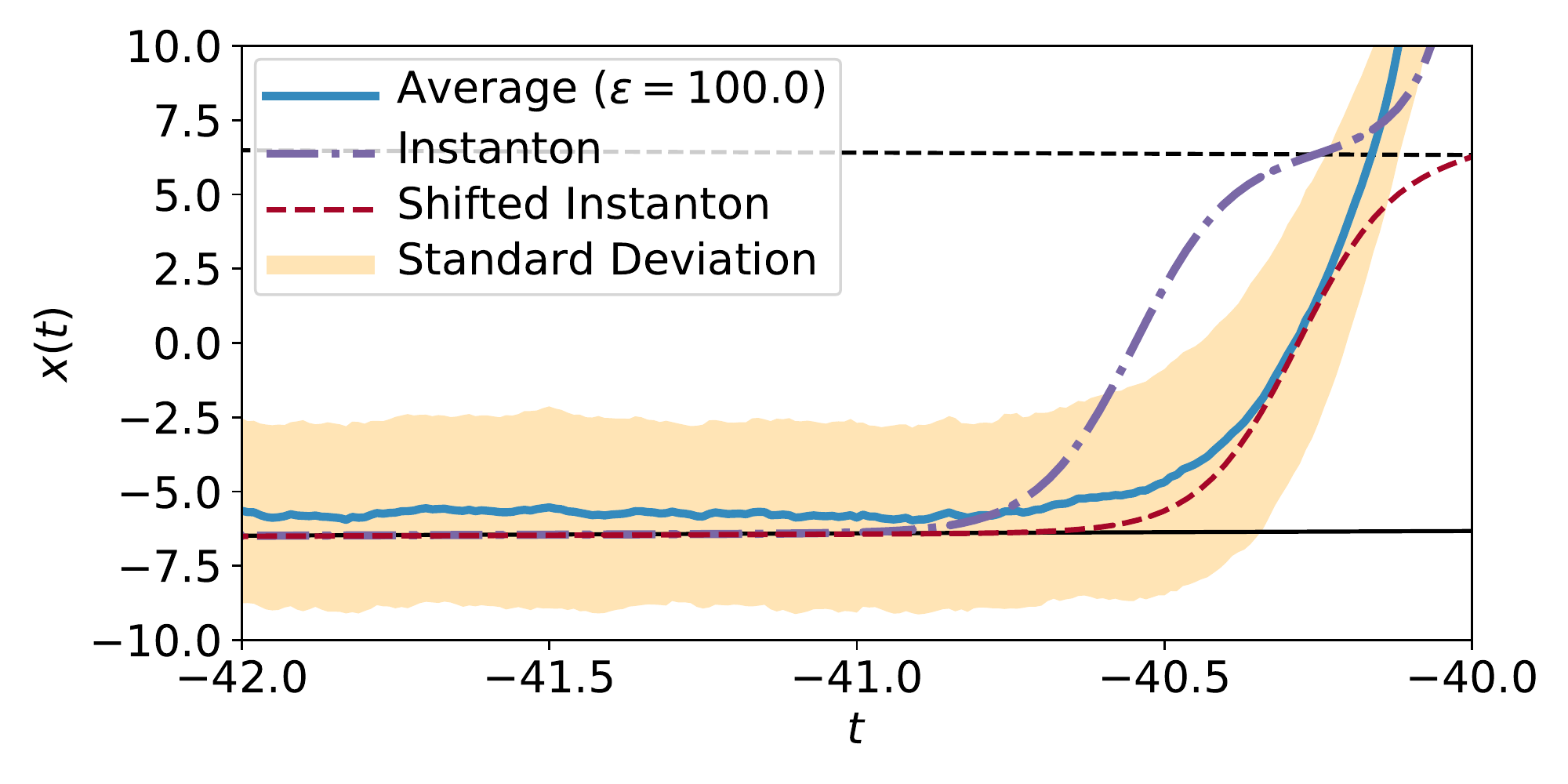}
  \caption{\label{fig:condtrajlarge} (Color online) Same as Fig.~\ref{fig:condtrajsmall}, for $\epsilon=100$. Conditioning on $\tau_M=-40$.}
\end{figure}
The situation is more complex in the large $\epsilon$ case.
In the classical case of time-independent potential barrier activation (e.g. the Kramers problem), the instanton is degenerated: it takes an infinite time to leave the attractor and an infinite time to reach the saddle-point~\cite{Caroli1981a}.
The time spent by reactive trajectories in the vicinity of the saddle-point is distributed according to a Gumbel law, and scales with $\ln \epsilon$.
This is why pure noise-induced transition times are unpredictable.
Here, it is very different (Fig.~\ref{fig:condtrajlarge}): the degeneracy of the instanton (starting from the attractor at $t\to-\infty$) is lifted because of the time dependence.
Nevertheless, compared to stochastic trajectories, the instanton triggers slightly before (shifting it in time describes correctly the dynamics away from the saddle-point) and still takes a longer time to pass the saddle-point.
This is because the amplitude of typical fluctuations is large enough to smooth the instanton slow-down at the saddle-point.
As a consequence, some trajectories remain in the vicinity of the saddle-point for a short while, but the majority of them swing directly to the other side of the potential.
This can also be seen as a consequence of the cost, in terms of the action functional, associated to staying at the position of the saddle-node, which is not a deterministic solution of the time-dependent problem.
Finally, let us note that because $\epsilon \gg 1$, we are no longer in the large deviation regime for $P[x]$, and there is \emph{a priori} no reason for the statistics of observables such as reacting trajectories to be dominated by an action-minimizing path.
This also holds when considering the appropriate action for finite $\epsilon$, $\mathcal{A}'=\mathcal{A}-\epsilon/2 \int dt \partial_x^2 V$~\cite{Graham1973}.
Hence, in the $\epsilon \gg 1$ regime, typical reacting trajectories are more predictable than the instanton in the sense that there is no imprint from the saddle-node, unlike the standard stationary case but similarly to the glacial-interglacial transitions (Fig.~\ref{fig:cycles}).
We have chosen a typical value for $\tau_M$ on Fig.~\ref{fig:condtrajlarge}, but the conclusions remain true for values which deviate significantly from the mean first-passage time.

\mysection{Conclusion}
We have given a global picture of the possible scenarios for transitions in a noisy system undergoing loss of stability, and the associated predictability.
We have shown that there exist two regimes characterized by a single control parameter $\epsilon$.
When $\epsilon$ is small, the escape time only deviates from the deterministic value in a Gaussian manner, and the reacting trajectories have a universal shape with an algebraic divergence.
On the contrary, when $\epsilon$ is large, escapes over the potential barrier become typical, but they are different from the standard Kramers problem: their \ac{pdf} is peaked, and can be predicted by an adiabatic approach consistent with large deviation theory.
Reacting trajectories leave the attractor exponentially fast and do not stick to the saddle-point.
Such trajectories are not described by large deviation theory.
These results open new prospects for the analysis of time series exhibiting abrupt transitions such as those encountered in climate dynamics.

\begin{acknowledgments}
  The authors would like to thank Mark Dykman and two anonymous referees for constructive comments which helped to improve the paper.
  The research leading to these results has received funding from the European Research Council under the European Union's seventh Framework Program (FP7/2007-2013 Grant Agreement No. 616811).
\end{acknowledgments}

\bibliographystyle{apsrev4-1}
\bibliography{bibtexlib}

\begin{thebibliography}{46}%
\makeatletter
\providecommand \@ifxundefined [1]{%
 \@ifx{#1\undefined}
}%
\providecommand \@ifnum [1]{%
 \ifnum #1\expandafter \@firstoftwo
 \else \expandafter \@secondoftwo
 \fi
}%
\providecommand \@ifx [1]{%
 \ifx #1\expandafter \@firstoftwo
 \else \expandafter \@secondoftwo
 \fi
}%
\providecommand \natexlab [1]{#1}%
\providecommand \enquote  [1]{``#1''}%
\providecommand \bibnamefont  [1]{#1}%
\providecommand \bibfnamefont [1]{#1}%
\providecommand \citenamefont [1]{#1}%
\providecommand \href@noop [0]{\@secondoftwo}%
\providecommand \href [0]{\begingroup \@sanitize@url \@href}%
\providecommand \@href[1]{\@@startlink{#1}\@@href}%
\providecommand \@@href[1]{\endgroup#1\@@endlink}%
\providecommand \@sanitize@url [0]{\catcode `\\12\catcode `\$12\catcode
  `\&12\catcode `\#12\catcode `\^12\catcode `\_12\catcode `\%12\relax}%
\providecommand \@@startlink[1]{}%
\providecommand \@@endlink[0]{}%
\providecommand \url  [0]{\begingroup\@sanitize@url \@url }%
\providecommand \@url [1]{\endgroup\@href {#1}{\urlprefix }}%
\providecommand \urlprefix  [0]{URL }%
\providecommand \Eprint [0]{\href }%
\providecommand \doibase [0]{http://dx.doi.org/}%
\providecommand \selectlanguage [0]{\@gobble}%
\providecommand \bibinfo  [0]{\@secondoftwo}%
\providecommand \bibfield  [0]{\@secondoftwo}%
\providecommand \translation [1]{[#1]}%
\providecommand \BibitemOpen [0]{}%
\providecommand \bibitemStop [0]{}%
\providecommand \bibitemNoStop [0]{.\EOS\space}%
\providecommand \EOS [0]{\spacefactor3000\relax}%
\providecommand \BibitemShut  [1]{\csname bibitem#1\endcsname}%
\let\auto@bib@innerbib\@empty
%</preamble>
\bibitem [{\citenamefont {Pierrehumbert}\ \emph {et~al.}(2011)\citenamefont
  {Pierrehumbert}, \citenamefont {Abbot}, \citenamefont {Voigt},\ and\
  \citenamefont {Koll}}]{Pierrehumbert2011}%
  \BibitemOpen
  \bibfield  {author} {\bibinfo {author} {\bibfnamefont {R.~T.}\ \bibnamefont
  {Pierrehumbert}}, \bibinfo {author} {\bibfnamefont {D.~S.}\ \bibnamefont
  {Abbot}}, \bibinfo {author} {\bibfnamefont {A.}~\bibnamefont {Voigt}}, \ and\
  \bibinfo {author} {\bibfnamefont {D.}~\bibnamefont {Koll}},\ }\href {\doibase
  10.1146/annurev-earth-040809-152447} {\bibfield  {journal} {\bibinfo
  {journal} {Ann. Rev. Earth Planet. Sci.}\ }\textbf {\bibinfo {volume} {39}},\
  \bibinfo {pages} {417} (\bibinfo {year} {2011})}\BibitemShut {NoStop}%
\bibitem [{\citenamefont {Paillard}(1998)}]{Paillard1998}%
  \BibitemOpen
  \bibfield  {author} {\bibinfo {author} {\bibfnamefont {D.}~\bibnamefont
  {Paillard}},\ }\href {\doibase 10.1038/34891} {\bibfield  {journal} {\bibinfo
   {journal} {Nature}\ }\textbf {\bibinfo {volume} {391}},\ \bibinfo {pages}
  {378} (\bibinfo {year} {1998})}\BibitemShut {NoStop}%
\bibitem [{\citenamefont {Huybers}\ and\ \citenamefont
  {Wunsch}(2005)}]{Huybers2005}%
  \BibitemOpen
  \bibfield  {author} {\bibinfo {author} {\bibfnamefont {P.}~\bibnamefont
  {Huybers}}\ and\ \bibinfo {author} {\bibfnamefont {C.}~\bibnamefont
  {Wunsch}},\ }\href {\doibase 10.1038/nature03401} {\bibfield  {journal}
  {\bibinfo  {journal} {Nature}\ }\textbf {\bibinfo {volume} {434}},\ \bibinfo
  {pages} {491} (\bibinfo {year} {2005})}\BibitemShut {NoStop}%
\bibitem [{\citenamefont {Crucifix}(2013)}]{Crucifix2013}%
  \BibitemOpen
  \bibfield  {author} {\bibinfo {author} {\bibfnamefont {M.}~\bibnamefont
  {Crucifix}},\ }\href {\doibase 10.5194/cp-9-2253-2013} {\bibfield  {journal}
  {\bibinfo  {journal} {Clim. Past}\ }\textbf {\bibinfo {volume} {9}},\
  \bibinfo {pages} {2253} (\bibinfo {year} {2013})}\BibitemShut {NoStop}%
\bibitem [{\citenamefont {Dansgaard}\ \emph {et~al.}(1993)\citenamefont
  {Dansgaard}, \citenamefont {Johnsen}, \citenamefont {Clausen}, \citenamefont
  {Dahl-Jensen}, \citenamefont {Gundestrup}, \citenamefont {Hammer},
  \citenamefont {Hvidberg}, \citenamefont {Steffensen}, \citenamefont
  {Sveinbj{\"o}rnsdottir}, \citenamefont {Jouzel},\ and\ \citenamefont
  {Bond}}]{Dansgaard1993}%
  \BibitemOpen
  \bibfield  {author} {\bibinfo {author} {\bibfnamefont {W.}~\bibnamefont
  {Dansgaard}}, \bibinfo {author} {\bibfnamefont {S.~J.}\ \bibnamefont
  {Johnsen}}, \bibinfo {author} {\bibfnamefont {H.}~\bibnamefont {Clausen}},
  \bibinfo {author} {\bibfnamefont {D.}~\bibnamefont {Dahl-Jensen}}, \bibinfo
  {author} {\bibfnamefont {N.}~\bibnamefont {Gundestrup}}, \bibinfo {author}
  {\bibfnamefont {C.}~\bibnamefont {Hammer}}, \bibinfo {author} {\bibfnamefont
  {C.}~\bibnamefont {Hvidberg}}, \bibinfo {author} {\bibfnamefont
  {J.}~\bibnamefont {Steffensen}}, \bibinfo {author} {\bibfnamefont
  {A.}~\bibnamefont {Sveinbj{\"o}rnsdottir}}, \bibinfo {author} {\bibfnamefont
  {J.}~\bibnamefont {Jouzel}}, \ and\ \bibinfo {author} {\bibfnamefont {G.~C.}\
  \bibnamefont {Bond}},\ }\href {\doibase 10.1038/364218a0} {\bibfield
  {journal} {\bibinfo  {journal} {Nature}\ }\textbf {\bibinfo {volume} {364}},\
  \bibinfo {pages} {218} (\bibinfo {year} {1993})}\BibitemShut {NoStop}%
\bibitem [{\citenamefont {Ganopolski}\ and\ \citenamefont
  {Rahmstorf}(2002)}]{Ganopolski2002}%
  \BibitemOpen
  \bibfield  {author} {\bibinfo {author} {\bibfnamefont {A.}~\bibnamefont
  {Ganopolski}}\ and\ \bibinfo {author} {\bibfnamefont {S.}~\bibnamefont
  {Rahmstorf}},\ }\href {\doibase 10.1103/PhysRevLett.88.038501} {\bibfield
  {journal} {\bibinfo  {journal} {Phys. Rev. Lett.}\ }\textbf {\bibinfo
  {volume} {88}},\ \bibinfo {pages} {38501} (\bibinfo {year}
  {2002})}\BibitemShut {NoStop}%
\bibitem [{\citenamefont {Ditlevsen}\ \emph {et~al.}(2007)\citenamefont
  {Ditlevsen}, \citenamefont {Andersen},\ and\ \citenamefont
  {Svensson}}]{Ditlevsen2007}%
  \BibitemOpen
  \bibfield  {author} {\bibinfo {author} {\bibfnamefont {P.~D.}\ \bibnamefont
  {Ditlevsen}}, \bibinfo {author} {\bibfnamefont {K.}~\bibnamefont {Andersen}},
  \ and\ \bibinfo {author} {\bibfnamefont {A.}~\bibnamefont {Svensson}},\
  }\href {\doibase 10.5194/cp-3-129-2007} {\bibfield  {journal} {\bibinfo
  {journal} {Clim. Past}\ }\textbf {\bibinfo {volume} {3}},\ \bibinfo {pages}
  {129} (\bibinfo {year} {2007})}\BibitemShut {NoStop}%
\bibitem [{\citenamefont {Jung}\ \emph {et~al.}(1990)\citenamefont {Jung},
  \citenamefont {Gray}, \citenamefont {Roy},\ and\ \citenamefont
  {Mandel}}]{Jung1990}%
  \BibitemOpen
  \bibfield  {author} {\bibinfo {author} {\bibfnamefont {P.}~\bibnamefont
  {Jung}}, \bibinfo {author} {\bibfnamefont {G.}~\bibnamefont {Gray}}, \bibinfo
  {author} {\bibfnamefont {R.}~\bibnamefont {Roy}}, \ and\ \bibinfo {author}
  {\bibfnamefont {P.}~\bibnamefont {Mandel}},\ }\href {\doibase
  10.1103/PhysRevLett.65.1873} {\bibfield  {journal} {\bibinfo  {journal}
  {Phys. Rev. Lett.}\ }\textbf {\bibinfo {volume} {65}},\ \bibinfo {pages}
  {1873} (\bibinfo {year} {1990})}\BibitemShut {NoStop}%
\bibitem [{\citenamefont {Rao}\ \emph {et~al.}(1990)\citenamefont {Rao},
  \citenamefont {Krishnamurthy},\ and\ \citenamefont {Pandit}}]{Rao1990}%
  \BibitemOpen
  \bibfield  {author} {\bibinfo {author} {\bibfnamefont {M.}~\bibnamefont
  {Rao}}, \bibinfo {author} {\bibfnamefont {H.~R.}\ \bibnamefont
  {Krishnamurthy}}, \ and\ \bibinfo {author} {\bibfnamefont {R.}~\bibnamefont
  {Pandit}},\ }\href {\doibase 10.1103/PhysRevB.42.856} {\bibfield  {journal}
  {\bibinfo  {journal} {Phys. Rev. B}\ }\textbf {\bibinfo {volume} {42}},\
  \bibinfo {pages} {856} (\bibinfo {year} {1990})}\BibitemShut {NoStop}%
\bibitem [{\citenamefont {Lo}\ and\ \citenamefont {Pelcovits}(1990)}]{Lo1990}%
  \BibitemOpen
  \bibfield  {author} {\bibinfo {author} {\bibfnamefont {W.~S.}\ \bibnamefont
  {Lo}}\ and\ \bibinfo {author} {\bibfnamefont {R.~A.}\ \bibnamefont
  {Pelcovits}},\ }\href {\doibase 10.1103/PhysRevA.42.7471} {\bibfield
  {journal} {\bibinfo  {journal} {Phys. Rev. A}\ }\textbf {\bibinfo {volume}
  {42}},\ \bibinfo {pages} {7471} (\bibinfo {year} {1990})}\BibitemShut
  {NoStop}%
\bibitem [{\citenamefont {Rahmstorf}(2002)}]{Rahmstorf2002}%
  \BibitemOpen
  \bibfield  {author} {\bibinfo {author} {\bibfnamefont {S.}~\bibnamefont
  {Rahmstorf}},\ }\href {\doibase 10.1038/nature01090} {\bibfield  {journal}
  {\bibinfo  {journal} {Nature}\ }\textbf {\bibinfo {volume} {419}},\ \bibinfo
  {pages} {207} (\bibinfo {year} {2002})}\BibitemShut {NoStop}%
\bibitem [{\citenamefont {Dijkstra}\ and\ \citenamefont
  {Ghil}(2005)}]{Dijkstra2005}%
  \BibitemOpen
  \bibfield  {author} {\bibinfo {author} {\bibfnamefont {H.}~\bibnamefont
  {Dijkstra}}\ and\ \bibinfo {author} {\bibfnamefont {M.}~\bibnamefont
  {Ghil}},\ }\href {\doibase 10.1029/2002RG000122} {\bibfield  {journal}
  {\bibinfo  {journal} {Rev. Geophys}\ }\textbf {\bibinfo {volume} {43}},\
  \bibinfo {pages} {3002} (\bibinfo {year} {2005})}\BibitemShut {NoStop}%
\bibitem [{\citenamefont {Eisenman}\ and\ \citenamefont
  {Wettlaufer}(2009)}]{Eisenman2009}%
  \BibitemOpen
  \bibfield  {author} {\bibinfo {author} {\bibfnamefont {I.}~\bibnamefont
  {Eisenman}}\ and\ \bibinfo {author} {\bibfnamefont {J.~S.}\ \bibnamefont
  {Wettlaufer}},\ }\href {\doibase 10.1073/pnas.0806887106} {\bibfield
  {journal} {\bibinfo  {journal} {Proc. Natl. Acad. Sci. U.S.A.}\ }\textbf
  {\bibinfo {volume} {106}},\ \bibinfo {pages} {28} (\bibinfo {year}
  {2009})}\BibitemShut {NoStop}%
\bibitem [{\citenamefont {Rose}\ \emph {et~al.}(2013)\citenamefont {Rose},
  \citenamefont {Ferreira},\ and\ \citenamefont {Marshall}}]{Rose2013}%
  \BibitemOpen
  \bibfield  {author} {\bibinfo {author} {\bibfnamefont {B.~E.~J.}\
  \bibnamefont {Rose}}, \bibinfo {author} {\bibfnamefont {D.}~\bibnamefont
  {Ferreira}}, \ and\ \bibinfo {author} {\bibfnamefont {J.}~\bibnamefont
  {Marshall}},\ }\href {\doibase 10.1175/JCLI-D-12-00175.1} {\bibfield
  {journal} {\bibinfo  {journal} {J. Climate}\ }\textbf {\bibinfo {volume}
  {26}},\ \bibinfo {pages} {2862} (\bibinfo {year} {2013})}\BibitemShut
  {NoStop}%
\bibitem [{\citenamefont {Arrhenius}(1889)}]{Arrhenius1889}%
  \BibitemOpen
  \bibfield  {author} {\bibinfo {author} {\bibfnamefont {S.}~\bibnamefont
  {Arrhenius}},\ }\href {\doibase 10.1016/B978-0-08-012344-8.50005-2}
  {\bibfield  {journal} {\bibinfo  {journal} {Z. Phys. Chem.}\ }\textbf
  {\bibinfo {volume} {4}},\ \bibinfo {pages} {226} (\bibinfo {year}
  {1889})}\BibitemShut {NoStop}%
\bibitem [{\citenamefont {Eyring}(1935)}]{Eyring1935}%
  \BibitemOpen
  \bibfield  {author} {\bibinfo {author} {\bibfnamefont {H.}~\bibnamefont
  {Eyring}},\ }\href {\doibase 10.1063/1.1749604} {\bibfield  {journal}
  {\bibinfo  {journal} {J. Chem. Phys.}\ }\textbf {\bibinfo {volume} {3}},\
  \bibinfo {pages} {107} (\bibinfo {year} {1935})}\BibitemShut {NoStop}%
\bibitem [{\citenamefont {Kramers}(1940)}]{Kramers1940}%
  \BibitemOpen
  \bibfield  {author} {\bibinfo {author} {\bibfnamefont {H.~A.}\ \bibnamefont
  {Kramers}},\ }\href {\doibase 10.1016/S0031-8914(40)90098-2} {\bibfield
  {journal} {\bibinfo  {journal} {Physica}\ }\textbf {\bibinfo {volume} {7}},\
  \bibinfo {pages} {284} (\bibinfo {year} {1940})}\BibitemShut {NoStop}%
\bibitem [{\citenamefont {Calef}\ and\ \citenamefont
  {Deutch}(1983)}]{Calef1983}%
  \BibitemOpen
  \bibfield  {author} {\bibinfo {author} {\bibfnamefont {D.~F.}\ \bibnamefont
  {Calef}}\ and\ \bibinfo {author} {\bibfnamefont {J.~M.}\ \bibnamefont
  {Deutch}},\ }\href {\doibase 10.1146/annurev.pc.34.100183.002425} {\bibfield
  {journal} {\bibinfo  {journal} {Ann. Rev. Phys. Chem.}\ }\textbf {\bibinfo
  {volume} {34}},\ \bibinfo {pages} {493} (\bibinfo {year} {1983})}\BibitemShut
  {NoStop}%
\bibitem [{\citenamefont {Bourgin}(1929)}]{Bourgin1929}%
  \BibitemOpen
  \bibfield  {author} {\bibinfo {author} {\bibfnamefont {D.~G.}\ \bibnamefont
  {Bourgin}},\ }\href {\doibase 10.1073/pnas.15.4.357} {\bibfield  {journal}
  {\bibinfo  {journal} {Proc. Natl. Acad. Sci. U.S.A.}\ }\textbf {\bibinfo
  {volume} {15}},\ \bibinfo {pages} {357} (\bibinfo {year} {1929})}\BibitemShut
  {NoStop}%
\bibitem [{\citenamefont {Wigner}(1932)}]{Wigner1932}%
  \BibitemOpen
  \bibfield  {author} {\bibinfo {author} {\bibfnamefont {E.}~\bibnamefont
  {Wigner}},\ }\href {\doibase 10.1103/PhysRev.40.749} {\bibfield  {journal}
  {\bibinfo  {journal} {Phys. Rev.}\ }\textbf {\bibinfo {volume} {40}},\
  \bibinfo {pages} {0749} (\bibinfo {year} {1932})}\BibitemShut {NoStop}%
\bibitem [{\citenamefont {Ravelet}\ \emph {et~al.}(2004)\citenamefont
  {Ravelet}, \citenamefont {Mari{\'e}}, \citenamefont {Chiffaudel},\ and\
  \citenamefont {Daviaud}}]{Ravelet2004}%
  \BibitemOpen
  \bibfield  {author} {\bibinfo {author} {\bibfnamefont {F.}~\bibnamefont
  {Ravelet}}, \bibinfo {author} {\bibfnamefont {L.}~\bibnamefont {Mari{\'e}}},
  \bibinfo {author} {\bibfnamefont {A.}~\bibnamefont {Chiffaudel}}, \ and\
  \bibinfo {author} {\bibfnamefont {F.}~\bibnamefont {Daviaud}},\ }\href
  {\doibase 10.1103/PhysRevLett.93.164501} {\bibfield  {journal} {\bibinfo
  {journal} {Phys. Rev. Lett.}\ }\textbf {\bibinfo {volume} {93}},\ \bibinfo
  {pages} {164501} (\bibinfo {year} {2004})}\BibitemShut {NoStop}%
\bibitem [{\citenamefont {Bouchet}\ and\ \citenamefont
  {Simonnet}(2009)}]{Bouchet2009}%
  \BibitemOpen
  \bibfield  {author} {\bibinfo {author} {\bibfnamefont {F.}~\bibnamefont
  {Bouchet}}\ and\ \bibinfo {author} {\bibfnamefont {E.}~\bibnamefont
  {Simonnet}},\ }\href {\doibase 10.1103/PhysRevLett.102.094504} {\bibfield
  {journal} {\bibinfo  {journal} {Phys. Rev. Lett.}\ }\textbf {\bibinfo
  {volume} {102}},\ \bibinfo {pages} {94504} (\bibinfo {year}
  {2009})}\BibitemShut {NoStop}%
\bibitem [{\citenamefont {Berhanu}\ \emph {et~al.}(2007)\citenamefont
  {Berhanu}, \citenamefont {Monchaux}, \citenamefont {Fauve}, \citenamefont
  {Mordant}, \citenamefont {P{\'e}tr{\'e}lis}, \citenamefont {Chiffaudel},
  \citenamefont {Daviaud}, \citenamefont {Dubrulle}, \citenamefont {Mari{\'e}},
  \citenamefont {Ravelet}, \citenamefont {Bourgoin}, \citenamefont {Odier},
  \citenamefont {Pinton},\ and\ \citenamefont {Volk}}]{Berhanu2007}%
  \BibitemOpen
  \bibfield  {author} {\bibinfo {author} {\bibfnamefont {M.}~\bibnamefont
  {Berhanu}}, \bibinfo {author} {\bibfnamefont {R.}~\bibnamefont {Monchaux}},
  \bibinfo {author} {\bibfnamefont {S.}~\bibnamefont {Fauve}}, \bibinfo
  {author} {\bibfnamefont {N.}~\bibnamefont {Mordant}}, \bibinfo {author}
  {\bibfnamefont {F.}~\bibnamefont {P{\'e}tr{\'e}lis}}, \bibinfo {author}
  {\bibfnamefont {A.}~\bibnamefont {Chiffaudel}}, \bibinfo {author}
  {\bibfnamefont {F.}~\bibnamefont {Daviaud}}, \bibinfo {author} {\bibfnamefont
  {B.}~\bibnamefont {Dubrulle}}, \bibinfo {author} {\bibfnamefont
  {L.}~\bibnamefont {Mari{\'e}}}, \bibinfo {author} {\bibfnamefont
  {F.}~\bibnamefont {Ravelet}}, \bibinfo {author} {\bibfnamefont
  {M.}~\bibnamefont {Bourgoin}}, \bibinfo {author} {\bibfnamefont
  {P.}~\bibnamefont {Odier}}, \bibinfo {author} {\bibfnamefont {J.-F.}\
  \bibnamefont {Pinton}}, \ and\ \bibinfo {author} {\bibfnamefont
  {R.}~\bibnamefont {Volk}},\ }\href {\doibase 10.1209/0295-5075/77/59001}
  {\bibfield  {journal} {\bibinfo  {journal} {EPL}\ }\textbf {\bibinfo {volume}
  {77}},\ \bibinfo {pages} {59001} (\bibinfo {year} {2007})}\BibitemShut
  {NoStop}%
\bibitem [{\citenamefont {Kurkij{\"a}rvi}(1972)}]{Kurkijarvi1972}%
  \BibitemOpen
  \bibfield  {author} {\bibinfo {author} {\bibfnamefont {J.}~\bibnamefont
  {Kurkij{\"a}rvi}},\ }\href {\doibase 10.1103/PhysRevB.6.832} {\bibfield
  {journal} {\bibinfo  {journal} {Phys. Rev. B}\ }\textbf {\bibinfo {volume}
  {6}},\ \bibinfo {pages} {832} (\bibinfo {year} {1972})}\BibitemShut {NoStop}%
\bibitem [{\citenamefont {Dykman}\ and\ \citenamefont
  {Krivoglaz}(1980)}]{Dykman1980}%
  \BibitemOpen
  \bibfield  {author} {\bibinfo {author} {\bibfnamefont {M.~I.}\ \bibnamefont
  {Dykman}}\ and\ \bibinfo {author} {\bibfnamefont {M.~A.}\ \bibnamefont
  {Krivoglaz}},\ }\href
  {http://www.sciencedirect.com/science/article/pii/0378437180900102}
  {\bibfield  {journal} {\bibinfo  {journal} {Physica A}\ }\textbf {\bibinfo
  {volume} {104}},\ \bibinfo {pages} {480} (\bibinfo {year}
  {1980})}\BibitemShut {NoStop}%
\bibitem [{\citenamefont {Dykman}(2012)}]{Dykman2012}%
  \BibitemOpen
  \bibfield  {author} {\bibinfo {author} {\bibfnamefont {M.}~\bibnamefont
  {Dykman}},\ }in\ \href {\doibase 10.1093/acprof:oso/9780199691388.001.0001}
  {\emph {\bibinfo {booktitle} {{Fluctuating Nonlinear Oscillators: From
  Nanomechanics to Quantum Superconducting Circuits}}}},\ \bibinfo {editor}
  {edited by\ \bibinfo {editor} {\bibfnamefont {M.}~\bibnamefont {Dykman}}}\
  (\bibinfo  {publisher} {Oxford University Press},\ \bibinfo {year}
  {2012})\BibitemShut {NoStop}%
\bibitem [{\citenamefont {Scheffer}\ \emph {et~al.}(2009)\citenamefont
  {Scheffer}, \citenamefont {Bascompte}, \citenamefont {Brock}, \citenamefont
  {Brovkin}, \citenamefont {Carpenter}, \citenamefont {Dakos}, \citenamefont
  {Held}, \citenamefont {Nes}, \citenamefont {Rietkerk},\ and\ \citenamefont
  {Sugihara}}]{Scheffer2009}%
  \BibitemOpen
  \bibfield  {author} {\bibinfo {author} {\bibfnamefont {M.}~\bibnamefont
  {Scheffer}}, \bibinfo {author} {\bibfnamefont {J.}~\bibnamefont {Bascompte}},
  \bibinfo {author} {\bibfnamefont {W.~A.}\ \bibnamefont {Brock}}, \bibinfo
  {author} {\bibfnamefont {V.}~\bibnamefont {Brovkin}}, \bibinfo {author}
  {\bibfnamefont {S.~R.}\ \bibnamefont {Carpenter}}, \bibinfo {author}
  {\bibfnamefont {V.}~\bibnamefont {Dakos}}, \bibinfo {author} {\bibfnamefont
  {H.}~\bibnamefont {Held}}, \bibinfo {author} {\bibfnamefont {E.~V.}\
  \bibnamefont {Nes}}, \bibinfo {author} {\bibfnamefont {M.}~\bibnamefont
  {Rietkerk}}, \ and\ \bibinfo {author} {\bibfnamefont {G.}~\bibnamefont
  {Sugihara}},\ }\href {\doibase 10.1038/nature08227} {\bibfield  {journal}
  {\bibinfo  {journal} {Nature}\ }\textbf {\bibinfo {volume} {461}},\ \bibinfo
  {pages} {53} (\bibinfo {year} {2009})}\BibitemShut {NoStop}%
\bibitem [{\citenamefont {Ditlevsen}\ and\ \citenamefont
  {Johnsen}(2010)}]{Ditlevsen2010}%
  \BibitemOpen
  \bibfield  {author} {\bibinfo {author} {\bibfnamefont {P.~D.}\ \bibnamefont
  {Ditlevsen}}\ and\ \bibinfo {author} {\bibfnamefont {S.~J.}\ \bibnamefont
  {Johnsen}},\ }\href {\doibase 10.1029/2010GL044486} {\bibfield  {journal}
  {\bibinfo  {journal} {Geophys. Res. Lett.}\ }\textbf {\bibinfo {volume}
  {37}},\ \bibinfo {pages} {L19703} (\bibinfo {year} {2010})}\BibitemShut
  {NoStop}%
\bibitem [{\citenamefont {Thompson}\ and\ \citenamefont
  {Sieber}(2011)}]{Thompson2011}%
  \BibitemOpen
  \bibfield  {author} {\bibinfo {author} {\bibfnamefont {J.~M.}\ \bibnamefont
  {Thompson}}\ and\ \bibinfo {author} {\bibfnamefont {J.}~\bibnamefont
  {Sieber}},\ }\href {\doibase 10.1142/S0218127411028519} {\bibfield  {journal}
  {\bibinfo  {journal} {Int. J. Bif. Chaos}\ }\textbf {\bibinfo {volume}
  {21}},\ \bibinfo {pages} {399} (\bibinfo {year} {2011})}\BibitemShut
  {NoStop}%
\bibitem [{\citenamefont {H{\"a}nggi}\ \emph {et~al.}(1990)\citenamefont
  {H{\"a}nggi}, \citenamefont {Talkner},\ and\ \citenamefont
  {Borkovec}}]{Hanggi1990}%
  \BibitemOpen
  \bibfield  {author} {\bibinfo {author} {\bibfnamefont {P.}~\bibnamefont
  {H{\"a}nggi}}, \bibinfo {author} {\bibfnamefont {P.}~\bibnamefont {Talkner}},
  \ and\ \bibinfo {author} {\bibfnamefont {M.}~\bibnamefont {Borkovec}},\
  }\href {\doibase 10.1103/RevModPhys.62.251} {\bibfield  {journal} {\bibinfo
  {journal} {Rev. Mod. Phys.}\ }\textbf {\bibinfo {volume} {62}},\ \bibinfo
  {pages} {251} (\bibinfo {year} {1990})}\BibitemShut {NoStop}%
\bibitem [{\citenamefont {Freidlin}\ and\ \citenamefont
  {Wentzell}(1998)}]{FreidlinWentzellBook}%
  \BibitemOpen
  \bibfield  {author} {\bibinfo {author} {\bibfnamefont {M.~I.}\ \bibnamefont
  {Freidlin}}\ and\ \bibinfo {author} {\bibfnamefont {A.~D.}\ \bibnamefont
  {Wentzell}},\ }\href@noop {} {\emph {\bibinfo {title} {{Random Perturbations
  of Dynamical Systems}}}},\ \bibinfo {edition} {2nd}\ ed.\ (\bibinfo
  {publisher} {Springer, New-York},\ \bibinfo {year} {1998})\BibitemShut
  {NoStop}%
\bibitem [{\citenamefont {{North Greenland Ice Core Project
  members}}(2004)}]{NGRIP2004}%
  \BibitemOpen
  \bibfield  {author} {\bibinfo {author} {\bibnamefont {{North Greenland Ice
  Core Project members}}},\ }\href {\doibase 10.1038/nature02805} {\bibfield
  {journal} {\bibinfo  {journal} {Nature}\ }\textbf {\bibinfo {volume} {431}},\
  \bibinfo {pages} {147} (\bibinfo {year} {2004})}\BibitemShut {NoStop}%
\bibitem [{\citenamefont {Benzi}\ \emph {et~al.}(1981)\citenamefont {Benzi},
  \citenamefont {Sutera},\ and\ \citenamefont {Vulpiani}}]{Benzi1981}%
  \BibitemOpen
  \bibfield  {author} {\bibinfo {author} {\bibfnamefont {R.}~\bibnamefont
  {Benzi}}, \bibinfo {author} {\bibfnamefont {A.}~\bibnamefont {Sutera}}, \
  and\ \bibinfo {author} {\bibfnamefont {A.}~\bibnamefont {Vulpiani}},\ }\href
  {\doibase 10.1088/0305-4470/14/11/006} {\bibfield  {journal} {\bibinfo
  {journal} {J. Phys. A}\ }\textbf {\bibinfo {volume} {14}},\ \bibinfo {pages}
  {L453} (\bibinfo {year} {1981})}\BibitemShut {NoStop}%
\bibitem [{\citenamefont {Gammaitoni}\ \emph {et~al.}(1998)\citenamefont
  {Gammaitoni}, \citenamefont {H{\"a}nggi}, \citenamefont {Jung},\ and\
  \citenamefont {Marchesoni}}]{Gammaitoni1998}%
  \BibitemOpen
  \bibfield  {author} {\bibinfo {author} {\bibfnamefont {L.}~\bibnamefont
  {Gammaitoni}}, \bibinfo {author} {\bibfnamefont {P.}~\bibnamefont
  {H{\"a}nggi}}, \bibinfo {author} {\bibfnamefont {P.}~\bibnamefont {Jung}}, \
  and\ \bibinfo {author} {\bibfnamefont {F.}~\bibnamefont {Marchesoni}},\
  }\href {\doibase 10.1103/RevModPhys.70.223} {\bibfield  {journal} {\bibinfo
  {journal} {Rev. Mod. Phys.}\ }\textbf {\bibinfo {volume} {70}},\ \bibinfo
  {pages} {223} (\bibinfo {year} {1998})}\BibitemShut {NoStop}%
\bibitem [{\citenamefont {Berglund}\ and\ \citenamefont
  {Gentz}(2002)}]{Berglund2002}%
  \BibitemOpen
  \bibfield  {author} {\bibinfo {author} {\bibfnamefont {N.}~\bibnamefont
  {Berglund}}\ and\ \bibinfo {author} {\bibfnamefont {B.}~\bibnamefont
  {Gentz}},\ }\href
  {http://iopscience.iop.org/article/10.1088/0951-7715/15/3/305/meta}
  {\bibfield  {journal} {\bibinfo  {journal} {Nonlinearity}\ }\textbf {\bibinfo
  {volume} {15}},\ \bibinfo {pages} {605} (\bibinfo {year} {2002})}\BibitemShut
  {NoStop}%
\bibitem [{\citenamefont {Berglund}\ and\ \citenamefont
  {Gentz}(2006)}]{BerglundGenzBook}%
  \BibitemOpen
  \bibfield  {author} {\bibinfo {author} {\bibfnamefont {N.}~\bibnamefont
  {Berglund}}\ and\ \bibinfo {author} {\bibfnamefont {B.}~\bibnamefont
  {Gentz}},\ }\href {\doibase 10.1007/1-84628-186-5} {\emph {\bibinfo {title}
  {{Noise-Induced Phenomena in Slow-Fast Dynamical Systems: A Sample-Paths
  Approach}}}},\ Probability and Its Applications\ (\bibinfo  {publisher}
  {Springer},\ \bibinfo {year} {2006})\BibitemShut {NoStop}%
\bibitem [{\citenamefont {Kuehn}(2011)}]{Kuehn2011}%
  \BibitemOpen
  \bibfield  {author} {\bibinfo {author} {\bibfnamefont {C.}~\bibnamefont
  {Kuehn}},\ }\href {\doibase 10.1016/j.physd.2011.02.012} {\bibfield
  {journal} {\bibinfo  {journal} {Physica D}\ }\textbf {\bibinfo {volume}
  {240}},\ \bibinfo {pages} {1020} (\bibinfo {year} {2011})}\BibitemShut
  {NoStop}%
\bibitem [{\citenamefont {Miller}\ and\ \citenamefont
  {Shaw}(2012)}]{Miller2012}%
  \BibitemOpen
  \bibfield  {author} {\bibinfo {author} {\bibfnamefont {N.~J.}\ \bibnamefont
  {Miller}}\ and\ \bibinfo {author} {\bibfnamefont {S.~W.}\ \bibnamefont
  {Shaw}},\ }\href {\doibase 10.1103/PhysRevE.85.046202} {\bibfield  {journal}
  {\bibinfo  {journal} {Phys. Rev. E}\ }\textbf {\bibinfo {volume} {85}},\
  \bibinfo {pages} {046202} (\bibinfo {year} {2012})}\BibitemShut {NoStop}%
\bibitem [{\citenamefont {Gardiner}(2009)}]{GardinerBook}%
  \BibitemOpen
  \bibfield  {author} {\bibinfo {author} {\bibfnamefont {C.~W.}\ \bibnamefont
  {Gardiner}},\ }\href@noop {} {\emph {\bibinfo {title} {{Handbook of
  Stochastic Methods for physics, chemistry, and the natural sciences}}}},\
  \bibinfo {edition} {4th}\ ed.\ (\bibinfo  {publisher} {Springer, Berlin},\
  \bibinfo {year} {2009})\BibitemShut {NoStop}%
\bibitem [{\citenamefont {Risken}(1989)}]{RiskenBook}%
  \BibitemOpen
  \bibfield  {author} {\bibinfo {author} {\bibfnamefont {H.}~\bibnamefont
  {Risken}},\ }\href@noop {} {\emph {\bibinfo {title} {The {F}okker-{P}lanck
  {E}quation}}},\ \bibinfo {edition} {2nd}\ ed.\ (\bibinfo  {publisher}
  {Springer},\ \bibinfo {year} {1989})\BibitemShut {NoStop}%
\bibitem [{\citenamefont {Erd{\'e}lyi}(1956)}]{ErdelyiBook}%
  \BibitemOpen
  \bibfield  {author} {\bibinfo {author} {\bibfnamefont {A.}~\bibnamefont
  {Erd{\'e}lyi}},\ }\href@noop {} {\emph {\bibinfo {title} {{Asymptotic
  Expansions}}}}\ (\bibinfo  {publisher} {Dover},\ \bibinfo {year}
  {1956})\BibitemShut {NoStop}%
\bibitem [{\citenamefont {Onsager}\ and\ \citenamefont
  {Machlup}(1953)}]{Onsager1953}%
  \BibitemOpen
  \bibfield  {author} {\bibinfo {author} {\bibfnamefont {L.}~\bibnamefont
  {Onsager}}\ and\ \bibinfo {author} {\bibfnamefont {S.}~\bibnamefont
  {Machlup}},\ }\href {\doibase 10.1103/PhysRev.91.1505} {\bibfield  {journal}
  {\bibinfo  {journal} {Phys. Rev.}\ }\textbf {\bibinfo {volume} {91}},\
  \bibinfo {pages} {1505} (\bibinfo {year} {1953})}\BibitemShut {NoStop}%
\bibitem [{\citenamefont {Machlup}\ and\ \citenamefont
  {Onsager}(1953)}]{Machlup1953}%
  \BibitemOpen
  \bibfield  {author} {\bibinfo {author} {\bibfnamefont {S.}~\bibnamefont
  {Machlup}}\ and\ \bibinfo {author} {\bibfnamefont {L.}~\bibnamefont
  {Onsager}},\ }\href {\doibase 10.1103/PhysRev.91.1512} {\bibfield  {journal}
  {\bibinfo  {journal} {Phys. Rev.}\ }\textbf {\bibinfo {volume} {91}},\
  \bibinfo {pages} {1512} (\bibinfo {year} {1953})}\BibitemShut {NoStop}%
\bibitem [{\citenamefont {Berglund}(2013)}]{Berglund2013}%
  \BibitemOpen
  \bibfield  {author} {\bibinfo {author} {\bibfnamefont {N.}~\bibnamefont
  {Berglund}},\ }\href
  {http://gateway.webofknowledge.com/gateway/Gateway.cgi?GWVersion=2&SrcAuth=mekentosj&SrcApp=Papers&DestLinkType=FullRecord&DestApp=WOS&KeyUT=000330087700006}
  {\bibfield  {journal} {\bibinfo  {journal} {Markov Processes Relat. Fields}\
  }\textbf {\bibinfo {volume} {19}},\ \bibinfo {pages} {459} (\bibinfo {year}
  {2013})}\BibitemShut {NoStop}%
\bibitem [{\citenamefont {Caroli}\ \emph {et~al.}(1981)\citenamefont {Caroli},
  \citenamefont {Caroli},\ and\ \citenamefont {Roulet}}]{Caroli1981a}%
  \BibitemOpen
  \bibfield  {author} {\bibinfo {author} {\bibfnamefont {B.}~\bibnamefont
  {Caroli}}, \bibinfo {author} {\bibfnamefont {C.}~\bibnamefont {Caroli}}, \
  and\ \bibinfo {author} {\bibfnamefont {B.}~\bibnamefont {Roulet}},\ }\href
  {\doibase 10.1007/BF01106788} {\bibfield  {journal} {\bibinfo  {journal} {J.
  Stat. Phys.}\ }\textbf {\bibinfo {volume} {26}},\ \bibinfo {pages} {83}
  (\bibinfo {year} {1981})}\BibitemShut {NoStop}%
\bibitem [{\citenamefont {Graham}(1973)}]{Graham1973}%
  \BibitemOpen
  \bibfield  {author} {\bibinfo {author} {\bibfnamefont {R.}~\bibnamefont
  {Graham}},\ }in\ \href {\doibase 10.1007/978-3-662-40468-3_1} {\emph
  {\bibinfo {booktitle} {Springer Tracts in Modern Physics}}},\ \bibinfo
  {editor} {edited by\ \bibinfo {editor} {\bibfnamefont {G.}~\bibnamefont
  {H{\"o}hler}}}\ (\bibinfo  {publisher} {Springer Verlag},\ \bibinfo {year}
  {1973})\ Chap.~\bibinfo {chapter} {1}, pp.\ \bibinfo {pages}
  {1--97}\BibitemShut {NoStop}%
\end{thebibliography}%

\end{document}